\documentclass[a4paper,12pt]{JHEP3}

\usepackage{graphicx}
\usepackage[latin1]{inputenc}
\usepackage{amssymb}
\usepackage{amsmath}
\usepackage{epsfig}

\newcommand{\ie}{{\it i.e.}}
\newcommand{\eg}{{\it e.g.}}

\newcommand{\pythia}{{\sc Pythia}}
\newcommand{\herwig}{{\sc Herwig}}

\newcommand{\be}{\begin{equation}}
\newcommand{\ee}{\end{equation}}
\newcommand{\ba}{\begin{eqnarray}}
\newcommand{\ea}{\end{eqnarray}}
\newcommand{\bt}{\begin{tabular}}
\newcommand{\et}{\end{tabular}}
\newcommand{\bfig}{\begin{figure}}
\newcommand{\efig}{\end{figure}}

\newcommand{\bbar}{\bar b}
\newcommand{\tbar}{\bar t}
\newcommand{\pT}{p_{\perp}}
\newcommand{\pt}[1]{p_{\perp, #1}}
\newcommand{\mt}[1]{m_{\perp, #1}}
\newcommand{\abs}[1]{\left|#1\right|}

\newcommand{\bigfrac}{\displaystyle\frac}
\newcommand{\smallfrac}{\textstyle\frac}

\newcommand{\GeV}{\mathrm{\;GeV}}

\preprint{TSL/ISV-2004-0282\\
September 2004}

\title{Improved description of 
charged Higgs boson production at hadron colliders}
\date{\today}
\author{J.~Alwall\\ 
High Energy Physics, Uppsala University, Box 535, S-751 21 Uppsala, Sweden\\
E-mail: \email{Johan.Alwall@tsl.uu.se}}
\author{J.~Rathsman\footnote{Research supported by the Swedish Research Council}\\ 
High Energy Physics, Uppsala University, Box 535, S-751 21 Uppsala, Sweden\\
E-mail: \email{Johan.Rathsman@tsl.uu.se}}

\abstract{ 
We present a new method for matching the two twin-processes $gb\to
H^\pm t$ and $gg\to H^\pm tb$ in Monte Carlo event generators.  The
matching is done by defining a double-counting term, which is used to
generate events that are subtracted from the sum of these two
twin-processes.  In this way we get a smooth transition between the
collinear region of phase space, which is best described by $gb\to
H^\pm t$, and the hard region, which requires the use of the $gg\to
H^\pm tb$ process.  The resulting differential distributions show
large differences compared to both the $gb\to H^\pm t$ and $gg\to
H^\pm tb$ processes illustrating the necessity to use matching
when tagging the accompanying $b$-jet.}

\begin{document}
\section{Introduction}

The search for physics beyond the Standard Model is one of the main
objectives of the upcoming experiments at the CERN Large Hadron
Collider (LHC) as well as the currently running ones at the Fermilab
Tevatron. Of special interest is the scalar (Higgs) sector, which
presumably is responsible for electroweak symmetry breaking and
generation of particle masses. The discovery of a neutral scalar Higgs
boson ($h^0$) would be a big step in understanding electroweak
symmetry breaking. But at the same time it may be difficult to
decide whether it is the Standard Model Higgs boson or if it belongs
to for example a supersymmetric theory. However, the discovery of a
charged Higgs boson would be a very clear signal of physics beyond the
Standard Model, and would give valuable insight into the parameter
space of this physics, whether supersymmetry or some other two Higgs
doublet model.

Present model-independent limits on the charged Higgs boson mass from LEP 
are $m_{H^\pm}>78.6$ GeV (95\% CL)~\cite{LEPlim} whereas the limits from the
Fermilab Tevatron are close to $m_{H^\pm}\gtrsim 160\;(130)$ GeV
~\cite{TEVlimCDF,TEVlimD0} for large (small) $\tan\beta$ 
($\tan\beta\gtrsim 100$ and $\tan\beta\lesssim 1$ respectively).
As usual, $\tan\beta=\bigfrac{v_1}{v_2}$ is the ratio of the vacuum 
expectation values of the two Higgs doublets which
determines the coupling strength to the top and bottom quarks
(roughly $\lambda_{H^\pm tb}\sim m_b\tan\beta + m_t\cot\beta $).
For a recent review of the prospects of further 
charged Higgs boson searches at the Tevatron and
the LHC see~\cite{Roy:2004az}.


\begin{figure}[ht]
\begin{center}
\epsfig{file=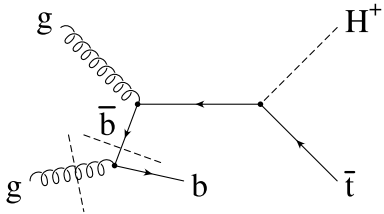,width=5cm}
\caption{\label{fig:overlap} Illustration of the relation between the  
$gb\to H^\pm t$ and $gg\to H^\pm tb$ processes.}
\end{center}
\end{figure}

At hadron colliders, the main contribution to direct single charged
Higgs boson production is through the two twin-processes $gb\to H^\pm
t$ and $gg\to H^\pm tb$~\footnote{For brevity we make no distinction
between quarks and anti-quarks unless it is not clear from
charge-conservation what the correct assignments are.}. The reason for
calling them twin-processes is simply that they describe the same
underlying physical process, as illustrated in fig.~\ref{fig:overlap},
but using different approximations. In fact, the latter process enters
in the next-to-leading order (NLO) calculation of the first
one~\cite{Zhu:2001nt,Plehn:2002vy,Berger:2003sm}. 
However, in the following we
will concentrate on the simulation of charged Higgs boson production
in Monte Carlo event generators such as \pythia~\cite{Pythia} and
\herwig~\cite{Herwig} where only tree-processes are
included\footnote{In the last years there has also been a development
of new techniques for including NLO matrix elements in Monte Carlo
generators, known as ``MC@NLO''~\cite{MCatNLO}. We will comment more on 
the relation of our method to MC@NLO below.}.
Note that although our matching
procedure works for all charged Higgs masses, we are mainly interested
in the region $m_{H^{\pm}}\gtrsim m_t-m_b$ where the
cross-sections for the two processes are of similar size.

In the $gb\to H^\pm t$ process the $b$-quark is considered
as a parton in the proton with a corresponding parton density which resums
all the, potentially large, 
leading logs of the type $(\alpha_s\log(\mu_F/m_b))^n$ that
arise when integrating the DGLAP evolution equation from the threshold
given by the $b$-quark mass to the factorization
scale $\mu_F$, where the proton is probed. Naturally, 
the treatment of the $b$-quark as a collinear massless parton in the proton 
leads to certain approximations which are at best only valid in the collinear 
limit. In contrast, the  $b$-quark is {\em not} considered as a massless 
parton in the $gg\to H^\pm tb$ process and consequently the kinematics of 
the process are exact. This also means that the finite parts, which are not
logarithmically enhanced, are also included correctly to order $\alpha_s^2$. 
On the other hand the $gg\to H^\pm tb$ process does {\em not} contain
the $(\alpha_s\log(\mu_F/m_b))^n$ terms with $n>1$ which are resummed in the 
$b$-quark density.

In case of the total cross-section 
it is straightforward how to combine the $gb\to H^\pm t$ 
and $gg\to H^\pm tb$ processes without introducing double 
counting~\cite{Borzumati:1999th} 
(see also~\cite{Barnett:1987jw,Olness:1987ep}). 
One simply adds the cross-sections from the two processes and subtracts
the common term, which is given by the $gb^\prime\to H^\pm t$ process 
where only the leading order logarithmically enhanced
 contribution to the $b$-quark density, 
$ b^\prime(x,\mu_F) \propto \alpha_s \log(\mu_F/m_b)$ is used. 
In this paper we will show how this
subtraction procedure can be extended also to the differential cross-section
(irrespective of whether the outgoing $b$-quark is
observed or not). One has to keep in mind
that even though the $b$-quark is considered to be collinear in the
$gb\to H^\pm t$ process and thereby the accompanying $b$-quark from 
the gluon splitting is also collinear, this is no longer true in a
Monte Carlo event generator. The reason is that in the Monte Carlo, 
the hard process is combined with
an initial state parton shower which ``undoes" the DGLAP evolution
back to a starting scale $Q_0 < m_b$, and thereby generates an
accompanying $b$-quark with non-zero $\pT$ (transverse momentum with
respect to the beam axis).

The experimental techniques for detecting charged Higgs bosons rely
heavily on tagging of $b$-jets and/or hadronic $\tau$-jets together
with missing $\pT$. The $b$-jets can originate from the charged Higgs
boson decay, the top quark decay or from an accompanying $b$-quark
whereas the $\tau$-jets and missing $\pT$ of interest come from the
charged Higgs boson decay.  Below the top quark mass the dominating
charged Higgs boson decay is $H^\pm \to \tau \nu_\tau $ whereas for
heavier masses the $H\to t b$ decay mode dominates. Thus the minimum
requirement is to tag one $b$-jet and either a $\tau$-jet together
with missing $\pT$ or two additional $b$-jets depending on the decay
mode of the charged Higgs boson. In addition the accompanying
$b$-quark may also be tagged.  Typical cuts used in studies by the
ATLAS~\cite{Assamagan:2002in,Assamagan:2002ne} and CMS~\cite{Kinnunen}
collaborations for $b$-jets, $\tau$-jets and missing $\pT$ are given
by the following: $\pt{b}>30$ GeV, $|\eta_b|<2.5$, $\pt{\tau}>100$
GeV, $|\eta_\tau|<2.5$, and $\pt{\mathrm{miss}}>100$ GeV. With these
cuts the $b$-tagging efficiency is of the order of $50$\% with a
miss-tagging rate of about 1\% whereas the $\tau$-tagging efficiency
is of the order of $40$\%~\cite{Kinnunen}.  In addition the difference
in polarization between a $\tau$-lepton coming from the scalar Higgs
decay and the vector $W$-boson decay leads to a harder spectrum when
the $\tau$ decays into one charged pion which in turn can be used to
enhance the signal~\cite{Roy:1999xw}.

Conventionally the strategy,
when investigating the prospects of detecting charged Higgs bosons at the LHC,
has been to use the $gb\to H^\pm t$ process
combined with parton showers when the accompanying $b$-quark is {\em not}
tagged in the final state (see \eg~\cite{Assamagan:2002in,Assamagan:2002ne}) 
whereas the 
$gg\to H^\pm tb$  process has been used when the $b$-quark is tagged
(see \eg~\cite{Miller:1999bm} and \cite{Assamagan:2004tt}).
As we will show in this paper, 
the only way to get reliable predictions
in the latter case is to make a proper matching of the two processes, 
whereas in the first case the need for matching depends on the details of the
selection used in defining the signal and the precision that one is aiming for.
We will also see that in general it is not possible to simply
divide the phase-space into two different parts, based for example on the
$p_\perp$ of the accompanying $b$-quark (as in \eg~\cite{Belyaev:2002eq}),
with each of them only populated by one of the processes.

The method that we present is general. The actual implementation has
been done in \pythia\ although the same procedure could also 
be applied to \herwig.

The outline of the rest of the paper is as follows. We start by recalling the
proper matching procedure in case of the total cross-section. In 
section 3 we then show how to generalize the method to be applicable 
also for differential cross-sections. The results of the matching
procedure are presented in section 4 and in section 5 we discuss how to choose
a proper factorization scale. Finally, in section 6 we give our conclusions and
a short outlook.

\section{The total cross-section}
As already discussed in the introduction, the two different approximations
describing $H^\pm$ production at hadron colliders that we are interested in 
are,
\ba
gb(\bbar) &\to& t H^- \; (\tbar H^+) \label{eq:LO} \\
gg   &\to& t \bbar H^- \;(\tbar b H^+) \label{eq:2to3} \quad .
\ea
The first one (\ref{eq:LO}), which we will denote the leading order (LO) or
$2\to2$ process,
includes the logarithmic DGLAP
resummation of gluon splitting to $b\bbar$ pairs via the $b$-quark
density, $b(\mu_F^2)\sim \sum (\alpha_s\log(\mu_F/m_b))^n$, 
whereas the second one, which we will denote the $2\to3$ process,
retains the correct treatment of the accompanying $b$-quark to order
$\alpha_s^2$.

In case of the total cross-section, the two approximations can be combined by 
simply adding them and subtracting the common term. The double-counting term
which needs to be subtracted is, to leading logarithmic accuracy, 
given by~\cite{Borzumati:1999th}
\be \label{eq:DC}
\sigma_\mathrm{DC}=\int dx_1dx_2\left[g(x_1,\mu_F^2)b'(x_2,\mu_F^2)
\frac{d\hat{\sigma}_{2\to 2}}{dx_1dx_2}(x_1,x_2) 
+ x_1 \leftrightarrow x_2\right]
\ee
where $b^\prime(x, \mu_F^2)$ is the leading logarithmic contribution to the 
$b$-quark density,
\be \label{eq:bprime}
 b^\prime(x, \mu_F^2)=
 \frac{\alpha_s(\mu_R^2)}{2\pi}\log\frac{\mu_F^2}{m_b^2}\int
 \frac{dz}{z} P_{g\to q\bar q}(z) \; g\left(\frac{x}{z},\mu_F^2\right)
\ee
and $P_{g\to q\bar q}(z)=\frac{1}{2}\left[z^2+(1-z)^2\right]$ is the
splitting function.

\EPSFIGURE[ht]
{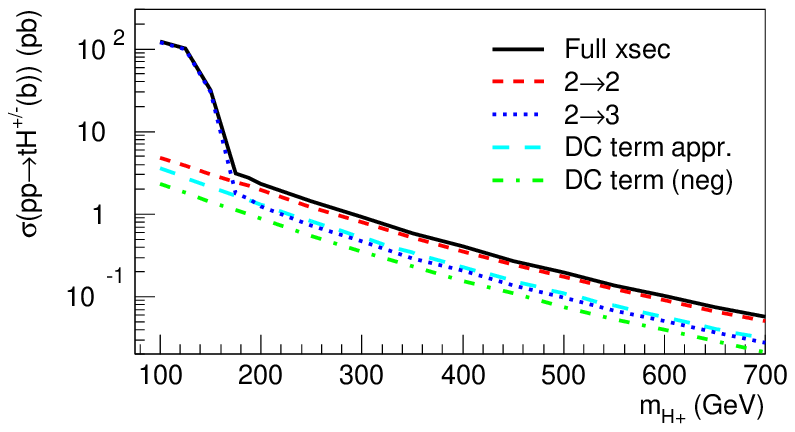,width=12cm} {\label{fig:massxsecs} Integrated
cross-section (components and matched total, using the exact
double-counting term) as a function of the $H^\pm$ mass at LHC. Here
$\tan\beta=30$ and $\mu_F=(m_t+m_{H^\pm})/4$. Note that the
double-counting term contribution (DC) is subtracted from the sum and
that the two different versions of the double-counting term
corresponds to whether the kinematic constraints are included or not.}

The matched integrated cross-section is thus given by
\be
\sigma = \sigma_{2\to 2} +\sigma_{2\to 3} \, -\sigma_\mathrm{DC}.
\label{eq:xsec}
\ee
The resulting cross-section for the process $pp \to H^\pm t b$ at the
LHC is illustrated in fig.~\ref{fig:massxsecs} as a function of
$m_{H^\pm}$ for the case $\tan\beta=30$ and using
$\mu_F=(m_t+m_{H^\pm})/4$.  (The choice of factorization scale will be
discussed in section \ref{sec:scales}.) We have used a running
$b$-quark mass in the Yukawa coupling evaluated at $m_{H^\pm}$, the
renormalization scale for $\alpha_s$ has also been set to $m_{H^\pm}$
and we have used the CTEQ5L~\cite{Lai:1999wy} parton densities. As can
be seen from fig.~\ref{fig:massxsecs}, for charged Higgs masses below
$m_t-m_b$ the $2\to3$ process dominates the cross-section. In this
region, the $2\to3$ process can be well approximated by intermediate
top production, $gg\to t\bar t\to tbH^\pm$. This approximation breaks
down for $m_{H^\pm}$ close to and above the top mass, see \eg\
\cite{Alwall:2003tc}. The matching procedure described in this paper, although it works for all charged 
Higgs masses, is of greatest interest in the regions where the $2\to2$
and $2\to3$ processes are of similar size, \ie\ for $m_{H^\pm}\gtrsim
m_t-m_b$.  At the same time our procedure gives a
smooth transition between the light and heavy Higgs mass regions,
which may be helpful when devising search strategies in this so called
transition region.

The way of defining the total cross-section in eq.~\eqref{eq:xsec} is
implicitly assuming that the accompanying $b$-quark from gluon
splitting is observed (which at least in principle is always possible
since there are no $b$-quarks in the initial state).  Thus we are in
effect talking about a leading order (in $\alpha_s$) cross-section for
the process $pp \to H^\pm t b$. Using the power counting rules of
ref.~\cite{Dicus:1998hs}, the first term in eq.~\eqref{eq:xsec},
$\sigma_{2\to 2}$, is of order $\alpha_s^2\log(\mu_F/m_b)$ whereas the
correction $(\sigma_{2\to 3} \, -\sigma_\mathrm{DC})$ is of order
$\alpha_s^2$. This $1/\log(\mu_F/m_b)$ correction, which arises from
gluon splitting into a $b\bar b$-pair, is also part of the NLO
calculation of the cross-section for the $pp \to H^\pm t $ process. In
addition the NLO calculation also includes virtual corrections as well
as real corrections from gluon emission both of which are of order
$\alpha_s^3\log(\mu_F/m_b)$. Another essential difference is of course
that in the NLO calculation the accompanying $b$-quark is not
observable.

Comparing with the MC@NLO approach as it is has been
implemented for heavy flavor 
production~\cite{Frixione:2003ei}, this uses the 
equivalent of $\sigma_{2\to 3}$ together with its NLO $\alpha_s$ corrections
as starting point. In other words, MC@NLO has so far only been
based on the so called 
flavor creation processes where there are no heavy quarks in the initial 
state and thus the equivalent of the $\sigma_{2\to 2}$ process
is not included. As a consequence the MC@NLO method has not yet been applied to
charged Higgs bosons production.

A more complete expression for $b^\prime$, which also takes into
account non-logarithmic contributions arising from kinematic
constraints due to the finite center of mass energy and the non-zero
$b$-quark mass, will be derived in the next section.  It is important
to include these corrections also in the double-counting term since
they are included both in the parton shower and the $2 \to 3$ matrix
element.  The difference when including the finite corrections are
substantial, leading to a large reduction of the double-counting term,
as can be seen from fig.~\ref{fig:massxsecs}.  For example, in the
case $m_{H^\pm}=250$ GeV, $\tan\beta=30$, and
$\mu_F=(m_t+m_{H^\pm})/4$ the double-counting term is reduced from
$0.81$ to $0.55$ pb.

For the purpose of the generalization of this method to the differential
cross-section, which will be given in the next section, it is instructive
to reconsider the derivation of the integrated double-counting term.
The starting point is the DGLAP evolution equation for a {\em massless} 
$b$-quark which we write,
\be \label{eq:DGLAP}
\frac{d b(x, k^2)}{d k^2}=
 \frac{1}{k^2}
 \frac{\alpha_s(\mu_R^2)}{2\pi} \int \frac{dz}{z} \left[
 P_{g\to q\bar q}(z) \; g\left(\frac{x}{z},k^2\right)
 +
 P_{q\to q g}(z) \; b\left(\frac{x}{z},k^2\right)
 \right]  .
\ee
Since we are only interested in the leading logarithmic contribution to
the $b$-quark density and it is generally assumed that the $b$-quark density
is only perturbatively generated\footnote{Thus we assume that the
intrinsic~\cite{Vogt:1994zf} contribution to the $b$-quark density can be neglected for
our purposes.} the last term can
be dropped. Furthermore, in the massive case, 
the $1/k^2$ term in (\ref{eq:DGLAP}), 
which originates from the $b$-quark propagator
(the quark line marked with a dash in fig.~\ref{fig:overlap}), 
has to be replaced with,
\be 
 \frac{1}{k^2} \to
 \frac{1}{k^2-m_b^2} =
 - \frac{1}{Q^2+m_b^2} 
\ee
where $Q^2=-k^2$. In summary we get the following expression,
\ba 
 b^\prime(x, \mu_F^2)
 &=&
 \frac{\alpha_s(\mu_R^2)}{2\pi}
 \int\frac{d Q^2}{Q^2+m_b^2}
 \int\frac{dz}{z} 
 P_{g\to q\bar q}(z) \; g\left(\frac{x}{z},Q^2\right) \\ 
 &\simeq&
 \frac{\alpha_s(\mu_R^2)}{2\pi}
 \int\frac{d Q^2}{Q^2+m_b^2}
 \int\frac{dz}{z} 
 P_{g\to q\bar q}(z) \; g\left(\frac{x}{z},\mu_F^2\right),
 \label{eq:bprimediff}
\ea
where the last step is valid to leading order in $\alpha_s$.
Eq.~(\ref{eq:bprime}) is then obtained by simply integrating
between the integration limits $Q^2_{\min}=m_b^2$  
and $Q^2_{\max}=\mu_F^2$,
which are correct to leading logarithmic accuracy.
In the next section we will derive the exact integration limits 
based on kinematic constraints 
that have to be taken into account in the differential cross-section.

\section{Matching the differential cross-section}
\label{sec:diff_matching}

In this section we give the details of how to extend the matching
procedure for the total cross-section outlined above to also be valid
for the differential cross-section. When doing this it is important to
keep in mind that in a Monte Carlo generator such as \pythia, the
leading order $2\to 2$ process is supplemented with initial state
parton showers which generate a non-zero $\pt{b}$ for the outgoing
$b$-quark from the gluon splitting, based on the leading order DGLAP
evolution equations. The $\pt{b}$-distribution generated in this way
is essentially of the type $d\sigma/d\pt{b} \propto \pt{b}/\mt{b}^2$,
where $\mt{b}^2=m_b^2+\pt{b}^2$ is the transverse mass squared, up to
the maximal $\pT$
generated by the parton shower, 
which in general\footnote{In \pythia\ the maximal
$\pT$ in the parton shower is by default equal to the factorization 
scale $\mu_F$ but it can also be set to be some factor times the 
factorization scale $\mu_F$.} should be given
by the factorization scale $\mu_F$.
This means that from the Monte Carlo point of view the final
state particles of the two processes (\ref{eq:LO}) and (\ref{eq:2to3})
are the same, but generated with different distributions due to the
different approximations used.

\begin{figure}[ht]
\begin{center}
\epsfig{file=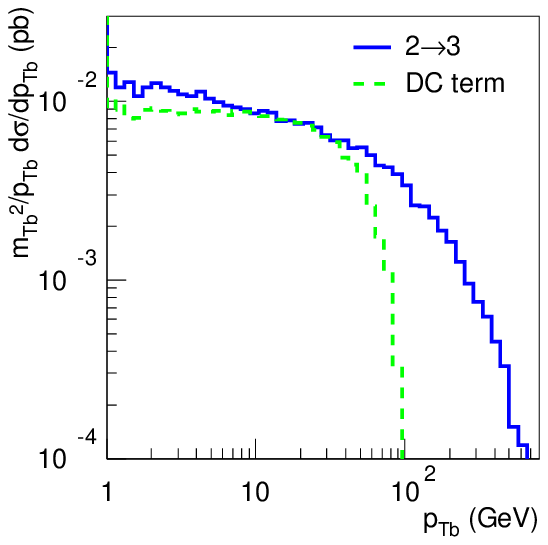,width=7cm}
\caption{\label{fig:plateau} The differential cross-section $d\sigma/d\pt{b}$
multiplied by $(m_b^2+\pt{b}^2)/\pt{b}$ for the $2\to 3$ matrix element and
the differential double-counting term, with $\mu_F=({m_t+m_{H^\pm}})/{4}$. 
Note the two distinct regions in the
$2\to 3$ cross-section, for small and large $\pt{b}$ respectively. }
\end{center}
\end{figure}

In order to distinguish the two processes and define the double-counting 
term we need some ``guiding principle'' which tells us in which region of phase
space the different approximations are valid. In our case it is natural
to use the $\pT$-distribution of the outgoing $b$-quark. To illustrate
the logic behind this choice it is instructive to consider the
$\pt{b}$-distribution from the matrix element
describing the $2\to3$ process, which is shown in
fig.~\ref{fig:plateau}. 
In order to get a better illustration of the behavior
at small $\pt{b}$ the differential cross-section has been multiplied by 
$\mt{b}^2/\pt{b}$. 

From the figure it is clear that for small $\pt{b}$ the differential
cross-section for the $2\to 3$ matrix element behaves more or less as
$d\sigma/d\pt{b} \propto \pt{b}/\mt{b}^2$ (similarly to the parton
shower), whereas for large $\pt{b}$ (where $\pt{b}$ is the 
hard scale of the process)
it behaves as $d\sigma/d\pt{b} \propto 1/\pt{b}^3$, as seen by
simple dimensional analysis.  This means that, as one expects on
general grounds, for small $\pt{b}$ the $2\to 3$ matrix element can be
factorized into two parts, the $g\to b\bar b$ splitting and the $gb
\to t H^\pm$ hard process (or symbolically $2\to3 = 1\to2 \otimes
2\to2$). For large $\pt{b}$ on the other hand, where the outgoing
$b$-quark is part of the hard process, it is not possible to factorize
the process and one has to retain the complete matrix element.

These observations can now be used to identify the regions of phase space
where the two approximations have their respective strength.
For small $\pt{b}$ we are in the collinear region where the cross-section
is dominated by logarithms of the type 
$\left(\alpha_s\log (\mt{b}^2/m_b^2)\right)^{n-1}/\mt{b}^2$ 
(the $2 \to 3$ matrix element only contains the leading term $n=1$) 
and therefore resummation effects, which are included in the
DGLAP evolution of the $b$-quark density, are important. For large 
$\pt{b}$ on the other hand we are in the hard region where factorization breaks
down and we have to use the $2 \to 3$ matrix element.

In order to obtain the correct differential cross-sections for the
accompanying $b$-quark, we need to employ a matching procedure between the
$gb\to H^\pm t$ and $gg\to H^\pm tb$ processes. Just as in the case of
the total cross-section the matching is done by adding the two
processes and subtracting the double-counting thus introduced in the
collinear region. The most basic requirements of such a procedure are
that all resulting differential cross-sections must be smooth, and
that the integrated cross-section must be the correct one as given by
eq.~(\ref{eq:xsec}).  Also the resulting differential cross-sections
should be given by the leading order process for small $\pt{b}$ and by
the $2\to 3$ matrix element for large $\pt{b}$.

The key observation in the generalization of the matching procedure to
the differential cross-section is that the leading logarithmic
$b$-quark density (\ref{eq:bprimediff}) defines a differential
distribution in the variables $z$ and $Q^2$, which together with the
differential cross-section for the $2\to2$ process and two azimuthal
angles gives a complete differential definition of the double-counting
term.  By picking events from this distribution and subtracting them
in the data analysis (\eg\ histograms), we can account for the
double-counting not only for the total cross-section but also
differentially.\footnote{The double-counting term is by construction
always smaller than the $2\to2$ term in the whole phase-space, since
$b'(x,Q^2) < b(x,Q^2)$. Therefore, if only the number of events is
large enough, there should be no risk that the matched cross-section
turns out to be negative in any phase-space bin.}

In this way, we can write down the complete expression for the
double-counting term based on eqs.~(\ref{eq:DC}) and
(\ref{eq:bprimediff}) (properly accounting for the phase-space
limitations as well as the mass of the incoming $b$-quark):

\ba
\lefteqn{\sigma_\mathrm{DC} = \int_{\tau_\mathrm{min}}^1\frac{d\tau}{\tau}
  \int_{\frac{1}{2}\log\tau}^{-\frac{1}{2}\log\tau}dy^*\frac{\pi}{\hat s}
  \int_{-1}^1\frac{\beta_{34}}{2}d(\cos\hat\theta)
  \left|\cal M_\mathrm{LO}\right|^2 \,\frac{\alpha_s(\mu_R^2)}{2\pi}\times}
  \nonumber \\
&& \left[\int_{x_1}^{z_\mathrm{max}}dz P_{g\to  q \bar q}(z)
 \int_{Q^2_\mathrm{min}}^{Q^2_\mathrm{max}}\frac{d(Q^2)}{Q^2+m_b^2} 
  \, \frac{x_1}{z} g(\frac{x_1}{z},\mu_F^2) \,x_2 g(x_2,\mu_F^2) 
  + x_1 \leftrightarrow x_2\right]
\label{eq:diff_dc}
\ea
Here $\cal M_\mathrm{LO}$ is the matrix element for the leading order
process (\ref{eq:LO}) and the variables for the
$2\to2$ process are defined as follows:
 $\tau= x_1 x_2$,
 $x_{1,2}=\sqrt{\tau}e^{\pm y^*}$,
 $\hat s = \tau s$, 
$\hat\theta$~is the polar angle of the $t$-quark in the CM system of 
 the $2\to 2$ scattering, and 
 $\beta_{34} = \hat s^{-1}\sqrt{(\hat s-m_t^2-m_{H^\pm}^2)^2-4m_t^2m_{H^\pm}^2}$.

The integration limits are given by
\begin{subequations}
\label{eq:limits}
\begin{align}
\tau_\mathrm{min}& = (m_t+m_{H^\pm})^2/s\\
z_\mathrm{max} &= \frac{Q^2_\mathrm{opt} \hat s}
		{(Q^2_\mathrm{opt}+\hat s)(Q^2_\mathrm{opt} + m_b^2)}
 \;,\;Q^2_\mathrm{opt}=\min\left(\sqrt{\hat s m_b^2},\mu_F^2\right)\\
Q^2_\mathrm{min}& = \smallfrac{1}{2}
		\left[\hat s\left(z^{-1}-1\right)-m_b^2\right] - 
	\smallfrac{1}{2}\sqrt{\left[\hat s\left(z^{-1}-1\right)-
		m_b^2\right]^2-4\hat sm_b^2}\\
Q^2_\mathrm{max}& = \min\left\{\mu_F^2,\smallfrac{1}{2}
		\left[\hat s\left(z^{-1}-1\right)-m_b^2\right] + 
	\smallfrac{1}{2}\sqrt{\left[\hat s\left(z^{-1}-1\right)-
		m_b^2\right]^2-4\hat sm_b^2}\right\}.
\end{align}
\end{subequations}
In deriving these limits we have identified $z$ as the ratio of the
center of mass energies of the $2\to2$ and $2\to3$ processes, $z=\hat
s/ \hat s^\prime$. Assuming that the outgoing $b$- and $t$-quarks as
well as the charged Higgs boson are on-shell, this identification of $z$
gives the following expression for the transverse momentum of the
outgoing $b$-quark in the center of mass system of the two gluons:
\be\label{eq:ptbrel}
\pt{b}^2 = Q^2-z\frac{(\hat s + Q^2)(m_b^2+Q^2)}{\hat s}.
\ee
Note that $Q^2=-k^2$, where $k$ is the 4-momentum of the $b$-quark
propagator. From this the limits on $z$ and $Q^2$ of
eq.~(\ref{eq:limits}) follow by considering the collinear situation
when $\pt{b}^2=0$ and taking into account that the virtuality
cannot be larger than the factorization scale $\mu_F^2$. In case of
the upper limit on $z$ one also has to find the optimal $Q^2$ value
which maximizes $z$ by setting $dz/dQ^2=0$.

It should be noted that the renormalization scale $\mu_R$ in
eq.~(\ref{eq:diff_dc}) is the same as in the hard $2\to2$ subprocess
(\ref{eq:LO}) and that the gluon density related to the $g\to b \bar
b$ splitting is evaluated at the same factorization scale $\mu_F$ as
the $b^\prime$-density and the other incoming gluon. These choices
corresponds to requiring that the double-counting term approximates
the $2 \to 3$ matrix element for small $\pt{b}$. In the \pythia\
parton shower the renormalization scale is given by
$(1-z)Q^2\simeq\pt{b}^2 $ and the factorization scale by $Q^2$, 
and since the parton shower corresponds to a
leading-logarithmic resummation, $\alpha_s$ is calculated at
1-loop. For $\pt{b} \sim
\mu_F$, the difference between the different scale choices is
negligible and thus the double-counting term will approach the
distribution obtained from the $2\to 2$ process with parton showers in
this region, ensuring that the matched differential cross-section
smoothly interpolates between the $2\to 2$ process for small $\pt{b}$
and the $2\to 3$ process for large $\pt{b}$.

Taking the limit $\hat s \to \infty$, eq.~(\ref{eq:ptbrel}) also gives 
the relation 
\be
\frac{d Q^2}{ Q^2 + m_b^2} = \frac{d \pt{b}^2}{ \pt{b}^2 + m_b^2}
\ee
which explains the plateau for small $\pt{b}$ seen in
fig.~\ref{fig:plateau} for both the double-counting term and the
$2\to3$ matrix element. The figure also illustrates the importance of
the kinematic constraints in the double-counting term which otherwise
would have been flat all the way up to $\pt{b}=\mu_F$. In
section~\ref{sec:scales} we will discuss how the extension of this
plateau can be used to set a proper
factorization scale for the $2\to 2$ process.

The differential cross-section for the double-counting term has been
implemented as an external process to \pythia\ and will be made
available for download~\cite{download} whereas the $2\to3$ process is
implemented as a regular process in \pythia\ from version 6.223.

\section{Results of the matching procedure}

In this section we present the results from a case study of
our matching procedure using $m_{H^\pm}=250\GeV$ and
$\tan\beta=30$ at LHC energies.

We use \pythia, including initial- and final-state radiation, with
default settings, except that the factorization scale is set to be
$\mu_F=(m_t+m_{H^\pm})/4$ (this choice of factorization scale will be
motivated in section~\ref{sec:scales}), and $\alpha_s$ in the hard process
and the running $b$-quark mass entering the Yukawa coupling are evaluated at
$m_{H^\pm}$ 
(we use a 1-loop expression for $\alpha_s$ and the Yukawa coupling 
since we are effectively dealing 
with a leading order process as argued in connection with eq.~\ref{eq:xsec}).
In addition the outgoing $b$-quark in the initial state
parton shower as well as in the double-counting term is put on-shell
in order to facilitate a more clearcut comparison with the $2\to 3$
process, where the outgoing $b$-quark is on-shell. In a realistic
case, using jet-finding algorithms, there should be no observable
difference compared to having the outgoing $b$-quark off-shell in the
parton shower.  Finally the maximal scale for the initial state shower is set
to the factorization scale and we use the CTEQ5L parton densities.

Since we are only dealing with leading order calculations the overall
normalization is quite uncertain, especially given that the
cross-section is proportional to the renormalized $b$-quark mass from
the Yukawa coupling. Therefore, in the following we will concentrate
on the shapes of the distributions.
The results for differential cross-sections in a few different
variables are presented in figs.~\ref{fig:diff-xsecs} 
and~\ref{fig:diff-xsecs2}.

\EPSFIGURE[ht]{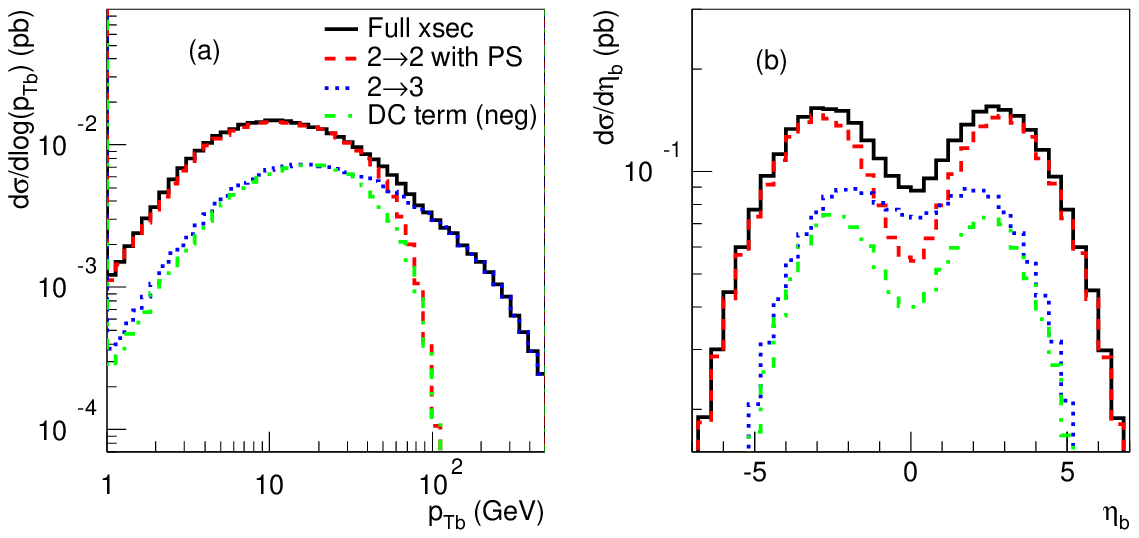,width=15cm}
{\label{fig:diff-xsecs} Differential distributions in 
(a) $\pt{b}$ and  (b) $\eta_b$ for the cross-section contributions
($2\to 2$ process, $2\to 3$ process and double-counting term) and
the resulting matched cross-section for $\tan\beta=30$,
$m_{H^\pm}=250\GeV$, $\mu_F=(m_t+m_{H^\pm})/4$. Note that the
double-counting term contribution (DC) is subtracted from the sum.}

First of all it is clear from fig.~\ref{fig:diff-xsecs}a
that the matched $\pt{b}$-distribution looks as expected. It follows the
leading order cross-section for small $\pt{b}$, where the $b$-quark is
collinear with the incoming gluon, and the $2\to 3$ process for large
$\pt{b}$, where the matrix element should give the correct $\pt{b}$
distribution. For intermediate $\pt{b} \lesssim \mu_F$, the double-counting 
term interpolates smoothly between the two approximations
giving an overall smooth transition between the collinear and hard
regions. We also note that the transition region is quite wide ranging
in this case from $\approx 10 \GeV$ to $\approx 100 \GeV$.

Restricting ourselves to the experimentally interesting region where
$b$-quarks can be tagged, \ie\ $\pt{b} \gtrsim 20 \GeV$ , we see that
the shape of the $\pt{b}$-distribution deviates substantially from the
one given by the $2\to3$ process. In fact, close to this lower limit
the matched distribution is about twice as large as the $2\to3$
process. In other words, if one only uses the $2\to3$ process, the
high-$\pt{b}$ region will be strongly overestimated compared to the
low-$\pt{b}$ one. 
Similarly for $\eta_b$
(fig.~\ref{fig:diff-xsecs}b), in the experimentally interesting
region $|\eta_b|<2.5$, the shape differs substantially both from the
leading order and the $2\to3$ processes, again illustrating that one has
to use matching in order to get a reliable description of the outgoing
$b$-quark.

From fig.~\ref{fig:diff-xsecs}a it is also clear that the method
suggested in~\cite{Belyaev:2002eq} for matching the LO and $2\to 3$
processes by making a cut in the $\pt{b}$, using the $2\to 3$ process
for events with $\pt{b}>\pt{\mathrm{cut}}$ and the LO process for
$\pt{b}<\pt{\mathrm{cut}}$, rescaled such that the total
cross-section is given by eq.~(\ref{eq:xsec}), does not work in
general. Such a matching procedure gives the correct cross-section
behavior for large $\pt{b}$ where both the leading order process and
the double-counting term are negligible, but the draw-back is that the
normalization is changed for small $\pt{b}$ relative to large $\pt{b}$
since the double-counting is only subtracted for $\pt{b} <
\pt{\mathrm{cut}}$. In addition the kinematic constraints represented
by the integration limits (\ref{eq:limits}) were not taken into account
leading to an underestimation of the cross-section, and it is also not
guaranteed that the differential cross-sections are smooth in all
kinematic variables.

In case one chooses not to tag the accompanying $b$-quark the
differences between the differential distributions from the $2\to 2$
process and the matched ones are not as significant. As an example we
have chosen to look at the distributions of the $b$-quark originating
from the top quark decay and the hadronic $\tau$-jet originating from
the charged Higgs bosons decay, which are of primary interest when
studying hadronic $\tau$-decays of heavy charged Higgs bosons (see
\eg~\cite{Assamagan:2002in}).

\EPSFIGURE[ht]{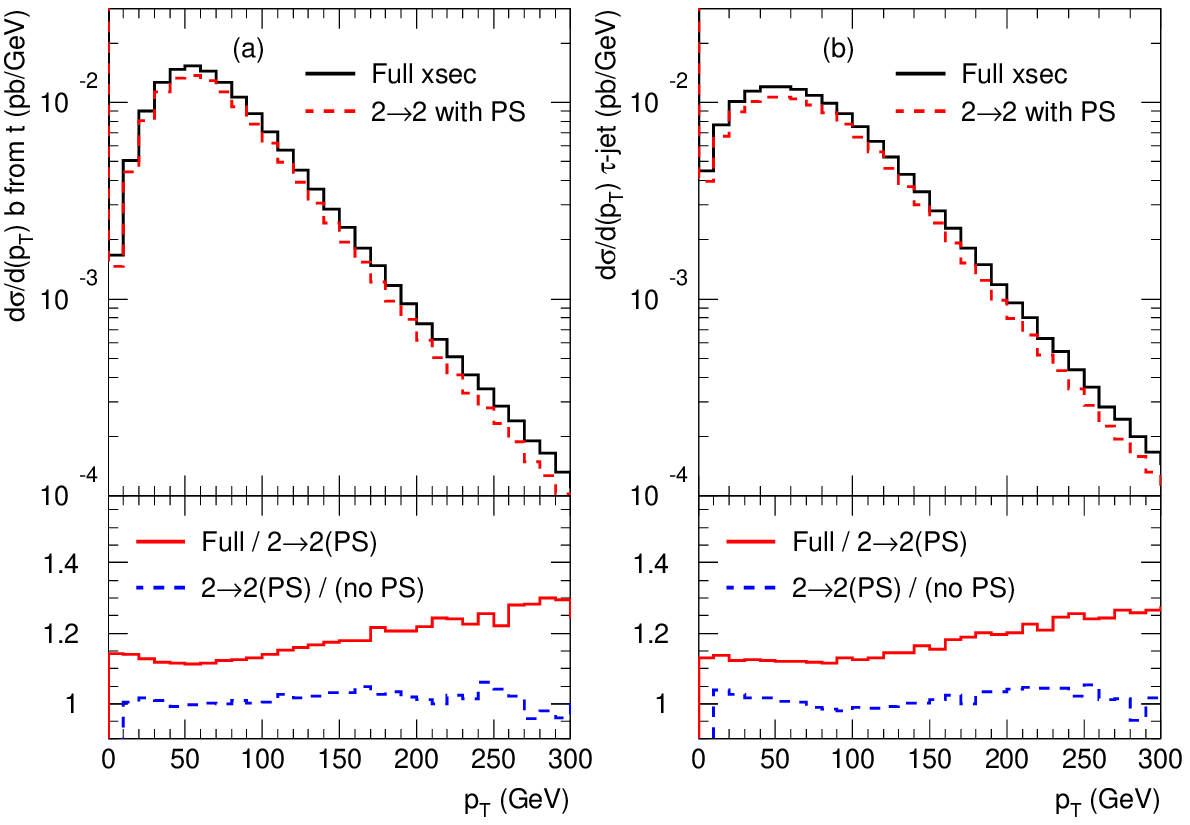,width=15cm}
{\label{fig:diff-xsecs2} Transverse momentum distributions of:
(a) the $b$-quark from the top decay, (b) the hadronic
$\tau$-jet from $H^\pm\to\tau^\pm\nu_\tau$, for $\tan\beta=30$,
$m_{H^\pm}=250\GeV$, and $\mu_F=(m_t+m_{H^\pm})/4$. The upper panels show
the matched result (solid) compared to the $2\to2$ process with parton 
showers (dashed), whereas the lower panels show the ratio of the two (solid) 
as well the ratio of the $2\to2$ process with and without parton showers (dashed). }

The effects of matching on these distributions are shown 
in fig.~\ref{fig:diff-xsecs2}. 
As can be seen from fig.~\ref{fig:diff-xsecs2}a 
the effects on the shape of the
transverse momentum distribution of the $b$-quark from the
top quark decay amounts to $10-15\%$ compared to
the leading order process in the region $\pt{b}\lesssim
250\GeV$. Similar results are obtained for the $\tau$-jet from the
decay of the charged Higgs boson, as seen in
fig.~\ref{fig:diff-xsecs2}b. These results, except for the
normalization, are not very sensitive to the choice of factorization
scale. As a way to gauge the importance of these effects 
fig.~\ref{fig:diff-xsecs2} also shows the effect of turning off the 
partons showers on these distributions. From the figure it is clear that 
with the scale for the partons showers set to $\mu_F=(m_t+m_{H^\pm})/4$, the
effects of matching are larger than the effects of the parton shower. However,
if one instead sets the scale of the parton shower to be $(m_t+m_{H^\pm})$, 
then the parton shower has a somewhat larger impact on the shape of the
distributions than the matching.

At the same time we have found that the effects of matching depend on
the cuts used in defining the signal. If the cuts are chosen carefully
as in~\cite{Assamagan:2002in} the differences in the shape of the
distributions are diminished.\footnote{The cuts used in this case are
in essence: one hadronic $\tau$ jet with $\pt{\tau}>100\GeV$ and
$\abs{\eta_\tau}\leq 2.5$, $\pT^{\rm miss}>100$ GeV, and at least one
$b$-tagged jet with $\pt{b}>30\GeV$ and $\abs{\eta_b}<2.5$. In
addition there is also anti-tag against additional $b$-jets by
requiring that there is not more than one $b$-jet with $\pt{b}>50\GeV$
and $\abs{\eta_b}<2$.}  As a consequence it is not possible to make a
general statement about the need to including matching when the
accompanying $b$-quark is not observed. Instead one has to investigate
this from case to case if one wants to pin down uncertainties of the
size illustrated in fig.~\ref{fig:diff-xsecs2}.

\section{Factorization scale dependence}
\label{sec:scales}

Until now we have used what may appear to be a rather small factorization scale
$\mu_F = (m_t+m_{H^\pm})/4$ compared to the more conventional
choice of $\mu_F \sim \sqrt{\hat{s}} \simeq m_t+m_{H^\pm}$. Of course,
in an all orders calculation the factorization scale dependence drops out,
but when the perturbative series is truncated one is left with a
residual scale dependence which can be quite large especially in leading order
calculations as we are dealing with here. However, this formal 
independence on the factorization scale when going to all orders does not 
mean that the choice is arbitrary. On the contrary, as we have already seen, 
the matrix element for the $2\to3$ process gives a clear indication
of where the transition between the collinear and hard regions takes place.
The important point to keep in mind is that the collinear parton density
integrates over all $\pT$ of the parton up to the factorization scale.
In other words, we can get a good indication of a proper choice of 
factorization scale for the leading order process from
the $\pt{b}$-distribution of the matrix element 
for the $2\to3$ process shown in fig.~\ref{fig:plateau}.

\EPSFIGURE[ht]{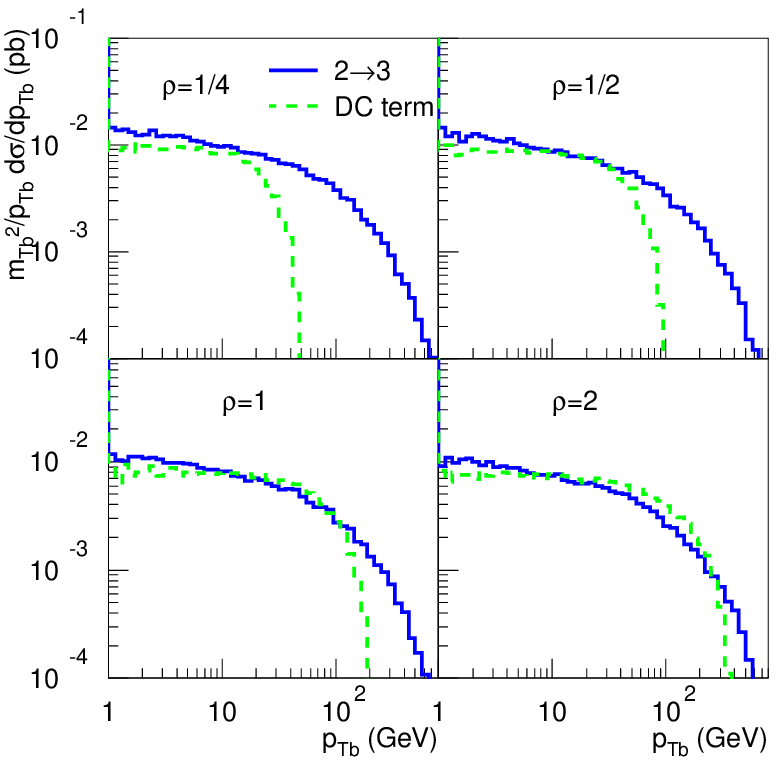,width=11cm}
{\label{fig:plateau4} The differential cross-section $d\sigma/d\pt{b}$
multiplied by $\mt{b}^2/\pt{b}$ for the $2\to 3$ matrix element and the
double-counting term for different factorization scales parameterized
by $\rho=2\mu_F/(m_t+m_{H^\pm})$.}

In order to be able to extract a suitable factorization scale from 
the $\pt{b}$-distribution it is instructive to compare the matrix element
with the double-counting term since the latter is based on the collinear 
approximation.
Naively one would expect the double-counting term
to have a flat $\pt{b}$-distribution when multiplied with $\mt{b}^2/\pt{b}$
all the way up to the factorization scale and then drop to zero. 
However, as is clear from 
fig.~\ref{fig:plateau} this is not true. The difference is mainly due to
the kinematic constraints and a non-constant gluon-density.
Comparing the two distributions for different factorization
scales, shown in fig.~\ref{fig:plateau4}, we see that
a suitable factorization scale is obtained by requiring that
the plateau in the distribution for the
double-counting term extends to the same $\pt{b}$ as the matrix element
but does not overshoot it. In this way we ensure that
the size of the collinear logarithms in the $b$-quark density
are not overestimated.
It follows that the factorization scale should be 
$\mu_F \approx (m_t+m_{H^\pm})/4$. 
We also see that the ``standard'' choice $\mu_F = m_t+m_{H^\pm}$ leads 
to a large overestimation of the collinear logs.

The $\pt{b}$-distribution was also used to find an appropriate factorization
scale in~\cite{Plehn:2002vy} and~\cite{Berger:2003sm} when
studying the  next-to-leading order corrections to the $gb \to tH^-$ process. 
In~\cite{Plehn:2002vy} the factorization scale was chosen such that the
integral over transverse momentum of the asymptotic form of the
differential cross-section, 
$\left.d\sigma/d\pt{b}\right|_{\rm asympt} = S \, \pt{b}/\mt{b}^2$,
gives the same total cross-section as the $2\to 3$ process,
whereas  in~\cite{Berger:2003sm} one compared the asymptotic behavior of the
transverse momentum distribution with the extent of the plateau in the $2\to 3$
process. Based on these arguments they suggest that a suitable factorization 
scale is 
$\mu_F \approx (m_t+m_{H^\pm})/6$ and $\mu_F \approx (m_t+m_{H^\pm})/5$
respectively which is very similar to what we find.

The question of finding an appropriate factorization scale is not particular to
charged Higgs boson production. In neutral Higgs boson production in 
association with bottom quarks one faces a similar situation where the question
of whether to treat the incoming $b$-quarks as partons or not is also important 
and one can use similar arguments for the appropriate factorization scale
to be used.
In~\cite{Rainwater:2002hm,Spira:2002rd} the transverse momentum distribution 
of $b$-quarks in the process $gg\to b \bar b h$ was compared to the
factorized expectation $d\sigma/d\pt{b} \propto \pt{b}/\mt{b}^2$. Using the
argument that these two distributions should approximately agree up to the
factorization scale, it was found that a factorization scale of order of 
$m_h/10$ is more proper than the usual $\mu_F={\cal O}(m_h)$. Based on the same
argument, a similar conclusion was also drawn in case of charged Higgs boson 
pair-production in association with bottom quarks~\cite{Moretti:2003px}.
In another approach~\cite{Maltoni:2003pn,Boos:2003yi} 
one instead looked at the distribution in Mandelstam-$t$ for the processes 
$b g \to h b$ and $b \bar b \to g h$ scaled with the Higgs boson mass, 
$-t d\sigma/dt ( \sqrt{-t}/m_h)$. Using these arguments the authors argue
that a small factorization scale $\mu_F \approx m_h/4$ is preferable.
Finally, comparing the NNLO calculation of 
$b\bar b \to h$ with the LO and NLO ones one~\cite{Harlander:2003ai} 
finds that both the scale dependence and higher order 
corrections are small when $(\mu_R,\mu_F) \sim (m_h,m_h/4)$.

\EPSFIGURE[ht]
{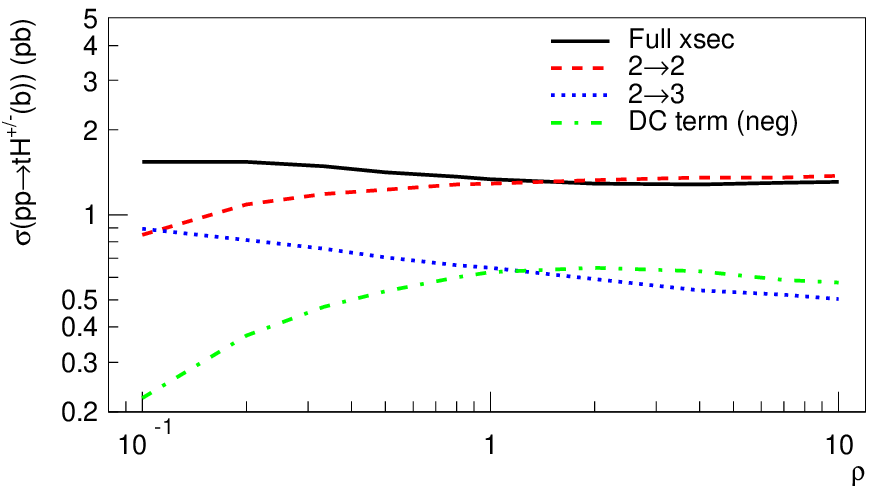,width=12cm}
{\label{fig:rhoxsecs} Integrated cross-sections at $m_{H^\pm}=250\GeV$ and
$\tan\beta=30$ as a 
function of the factorization scale parametrized by 
$\rho=2\mu_F/(m_t+m_{H^\pm})$.}

Given that the arguments for choosing the factorization scale to be
$\mu_F = (m_t+m_{H^\pm})/4$ are not very precise we have studied to what 
extent the results of the matching procedure changes when
varying the factorization scale (in most cases by varying it by a factor two
up or down). In order not to confuse the picture we have kept the
renormalization scale in $\alpha_s$ and the running $b$-quark mass fixed.

\EPSFIGURE[ht]
{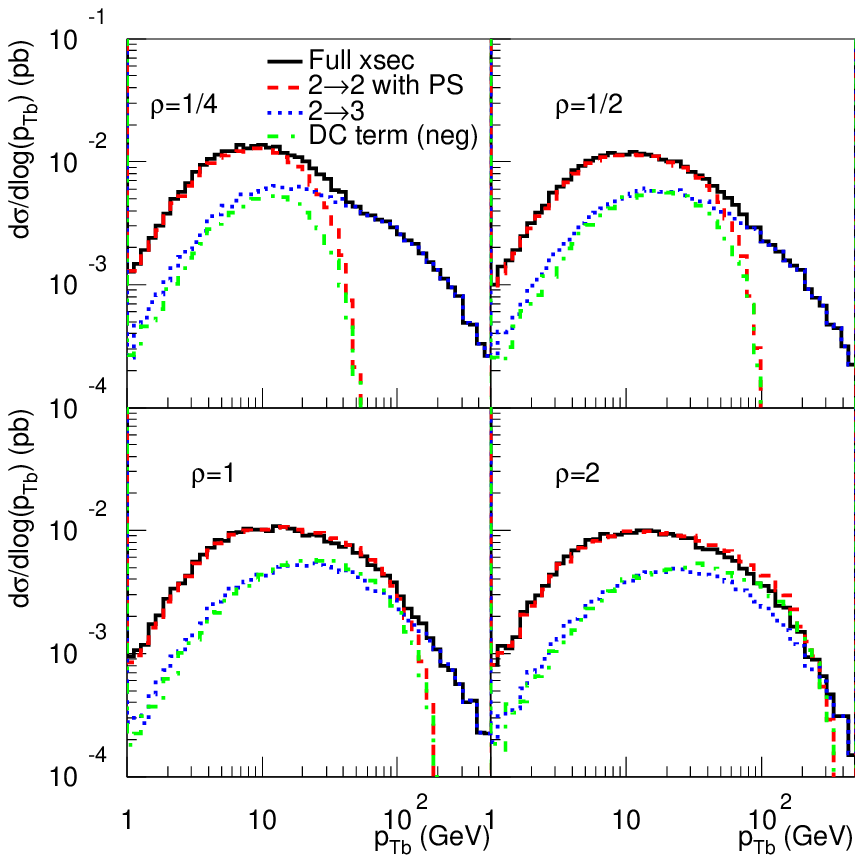,width=12cm} {\label{fig:ptxsecs} The different
components of the differential cross-section in $\pt{b}$ for different
choices of the factorization scale parametrized by
$\rho=\frac{2\mu_F}{(m_t+m_{H^\pm})}=1/4,\,1/2,\,1,\,2$.}

We start out by considering the factorization scale dependence of the matched
total cross-section. Parameterizing the factorization scale using the 
parameter $\rho$:
\be
\mu_F =\rho\,\frac{m_t+m_{H^\pm}}{2}
\ee
we get the result shown in fig.~\ref{fig:rhoxsecs}. Comparing with 
the scale dependence of the different components ($2\to2$, $2\to 3$ and 
double-counting)  which are also displayed we see that the scale dependence is
substantially  reduced after matching.
For example, varying the factorization scale a factor 2 up or down from 
$\rho=0.5$ the cross-section changes with less than 10\%.
This reduced factorization scale dependence 
is expected since we are including parts of the NLO corrections 
(more specifically the real corrections from gluon splitting into 
$b\bar b$-pairs in the initial state) to the leading order process. 
The same reduction of the scale dependence has also been seen in neutral Higgs
boson production via $b$-quark fusion~\cite{Maltoni:2003pn}.
In addition we see that at $\rho\gtrsim 1$ the double-counting
term becomes larger than the $2\to 3$ process. In view of the
fact that the double-counting term is included in order to cancel the
component of the $2\to 3$ process that is already
included in the leading order process, this  marks a
breakdown in the procedure, indicating that the factorization scale is
chosen too large.

Next we consider the matching procedure itself. As can be seen from
fig.~\ref{fig:ptxsecs}, showing the $\pT$-distribution of the $b$-quark, the
matching procedure still fulfills the requirements outlined  in
section~\ref{sec:diff_matching} for other factorization scales but the relative
importance of the different contributions varies. However, for the case $\rho
\gtrsim 1$, the double-counting term is overshooting the $2\to3$
process in the region  $\pt{b} \lesssim \mu_F$, again
indicating that this choice of factorization scale is too large.

\begin{figure}[t]
\begin{center}
\epsfig{file=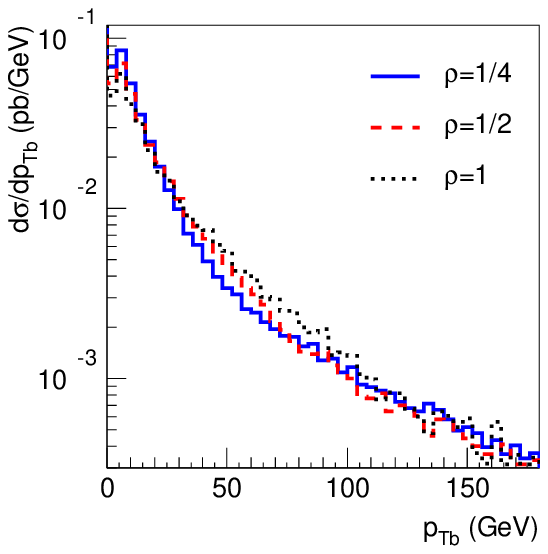,width=7cm}
\caption{\label{fig:sumxsecs} Matched $\pt{b}$-differential cross-sections for 
different choices of the factorization scale, 
$\rho=\frac{2\mu_F}{(m_t+m_{H^\pm})}=1/4,\,1/2,\,1$. }
\end{center}
\end{figure}


Finally fig.~\ref{fig:sumxsecs} shows the relative stability of the
matched $\pt{b}$-distribution when varying the factorization scale
up or down with a factor 2 from our preferred value. 
From the figure we see that
the residual factorization scale uncertainty is typically small and
at most $\sim20\%$ (in the
region $\pt{b} \sim (m_t+m_{H^\pm})/5 $). 
This should be compared with the difference
between the matched distribution and the one from the $2\to3$ matrix element
which amounts to a factor $\sim2$ at $\pt{b}\sim 20$ GeV.

\section{Conclusions and Outlook}

In this paper we have presented a new method for matching the $gb\to H^\pm t$
 and $gg\to H^\pm tb$ processes in Monte Carlo event generators such as
\pythia\ and \herwig. By matching the two processes at hand in a proper
way we can combine their respective virtues. On the one hand, the $gb\to H^\pm t$
(or $2\to2$) process includes a resummation of potentially large logarithms of the type
$\alpha_s\log(\mu_F/m_b)$ which arise from the collinear region of phase space
where the transverse momentum of the accompanying $b$-quark $\pt{b}$ is
small. On the other hand, the $gg\to H^\pm tb$ (or $2\to3$) processes contains the exact
kinematics of the accompanying $b$-quark which is important in the hard region
where $\pt{b}$ is large.

The matching is done by adding the 
two different approximations and subtracting the common part which
otherwise would be double-counted. In other words the double-counting term
is given by the collinear approximation of the $gg\to H^\pm tb$ process.
By viewing the double-counting term as a differential distribution in the
kinematic variables of the $2\to3$ process, it can be used to generate events 
in the same way as any other process, including parton showers and
hadronization. 
These events are then subtracted from the sum of the 
events generated from the $gb\to H^\pm t$ and $gg\to H^\pm tb$ processes.

This matching procedure leads to a smooth transition between the
collinear region of phase space, where the $2\to2$ process dominates,
and the hard region where the $2\to3$ process dominates. Looking at
the $\pt{b}$-distribution the difference between the matched
cross-section and the $2\to3$ process can be as large as a factor
$\sim2$ in regions of experimental interest ($\pt{b}\gtrsim 20$
GeV). Likewise, the pseudorapidity distribution of the accompanying
$b$-quark differs significantly in shape from both the $2\to2$ and the
$2\to3$ process in the central region. When looking at the
distributions of the decay products of the $t$-quark and the charged
Higgs boson the differences are smaller, typically $\sim 10\%$ in
experimentally interesting regions of phase space which is similar to
the effects of parton showers.  At the same time, the differences turn
out to be sensitive to the cuts used to define the signal and by
carefully choosing the cuts, as in~\cite{Assamagan:2002in}, the
effects can be made negligible.

Not only does the matching procedure give a better description of
charged Higgs boson production, in addition it also gives a reduced
factorization scale dependence. The sum of the different contributions
to the total cross-section has a much smaller factorization scale
dependence than the individual parts. Looking at the
$\pt{b}$-distribution from the $gg\to H^\pm tb$ matrix element and
comparing to the double-counting term we also get strong arguments for
choosing a factorization scale $\mu_F = (m_t+m_{H^\pm})/4$ which is
substantially smaller than the ``standard choice" $\mu_F \sim
\sqrt{\hat{s}} \simeq (m_t+m_{H^\pm})$.  Being conservative we
estimate the remaining factorization scale dependence by varying the
factorization scale with a factor two. Doing this we find that the
total cross-section varies with about $\pm10\%$ and that the height of the
$\pt{b}$-distribution is also quite stable, varying
at most about $\pm20\%$.

The method that we have presented in this paper is not restricted to charged
Higgs boson production. It can also be applied in other processes where one has
incoming $b$-quarks. The simplest example is $gb\to W^\pm t$-production, but
it can also be extended to include processes with two incoming $b$-quarks,
such as Higgs boson production in association with $b$-quarks. 
It would also be
interesting to extend the method to also include the remaining NLO corrections
to get a NLO normalization of the total cross-section as in the MC@NLO method.

\begin{acknowledgments}
We are grateful to Nils Gollub, Gunnar Ingelman, and Stefano Moretti for 
comments on the manuscript.
\end{acknowledgments}

\end{document}